\documentclass[12pt]{article} \usepackage{times}
\usepackage{graphicx} 
\begin{document} \title{Four classes of modified relativistic 
symmetry
transformations} \author{Jerzy Lukierski \\ 
Theoretical Physics, \\
  University of Wroc{\l}aw, pl. 
Maxa Borna 9, 
 \\
50-204 Wroc{\l}aw, Poland
\\ \\
Anatol Nowicki\thanks{Talk given by A. Nowicki}\ 
\thanks{Supported 
by the Polish State Committee for Scientific Research 
 (KBN) grant No 5P03B05620}
\\ 
Institute of Physics, University of Zielona G\'{o}ra, 
\\ 
 ul. Podg\'{o}rna 50, 65-246 
 Zielona G\'{o}ra, Poland}

\date{}     
\maketitle

\begin{abstract}
We discuss the nonlinear transformations of standard Poincar\'{e}
symmetry in   the context of recently introduced Doubly Special
Relativity (DSR) theories. We introduce four classes of modified
relativistic theories with three of them describing various DSR
frameworks. We consider four examples of modified relativistic
symmetries, which illustrate each of the considered class.
\end{abstract}

\section{Introduction}

The classical relativistic symmetries as described by classical
Poincare-Hopf algebra can be deformed due to the following two
reasons (see also [1,2]):
\begin{description}
\item{ i)} One can introduce new nonlinear basis in enveloping
algebra of classical Poincar\'{e} Lie algebra,
\item{ii)} One can deform the classical coalgebraic structure,
which leads to quantum Poincar\'{e} algebras and by considering
dual Hopf algebra structure also provides the quantum Poincar\'{e}
groups.
\end{description}

In this talk we shall consider the modification of relativistic
symmetry transformations due to the nonlinear change of basis in
the algebraic sector. Recently Amelino-Camelia [3] introduced two
categories of modified relativistic symmetries with two parameters
$c$ and $\kappa$\\-- light velocity $c$ and fundamental mass
$\kappa$ which can be identified with Planck mass -- invariant
under the modified Lorentz transformations:
\begin{description}
\item{i)} Doubly special relativistic theories of first type, denoted by
DSR1, with energy $E$ unbounded and momentum $\vec{P}$ 
bounded by
mass-like parameter $\kappa$:  $|\vec{P}| \leq \kappa \, c$ \item{ii)}
DSR2 theories, with both energy and moment bounded by 
$\kappa$: $|\vec{P}|
\leq \kappa \, c$ and   $ E \leq \kappa \, c^2.$ The first examples of
DSR2 theories were provided by Magueijo and Smolin [4,5].
\end{description} In  order to complete the classification we shall
introduce further two categories:

\begin{description}
\item{iii)}  DSR3 theories, with momentum unbounded and energy
bounded: $E \leq \kappa c^2$.
\item{iv)} The theories with momentum as well as energy 
unbounded. In such
a framework two parameters $c$ and $\kappa$ do not have the 
meaning of
invariant parameters. We shall call these theories Smoothly 
Modified
Special Relativity (SMSR) theories. \end{description}

\section{Nonlinear realizations of relativistic symmetries}

We use the following notation in the description of classical D=4
Poincar\'{e} algebra:\\-- Lorentz algebra
($g_{\mu\nu}=(-1,1,1,1)$)
\begin{equation}\label{pary2}
[M_{\mu\nu},M_{\rho\tau}] = i(g_{\mu\rho}M_{\nu\tau}+
g_{\nu\tau}M_{\mu\rho}-g_{\mu\tau}M_{\nu\rho}
-g_{\nu\rho}M_{\mu\tau}).
\end{equation}
-- covariance relations  ($M_i= \frac12\epsilon_{ijk} M_{jk},
N_i=M_{i0}$; $c$ is the light velocity)
\begin{eqnarray}
[M_i, {\cal{P}}_j] =  i\epsilon_{ijk}{\cal{P}}_k ,\quad [M_i,
{\cal{E}} ] = 0 , \quad [N_i,{\cal{P}}_j]& =  {i\over
c}\delta_{ij}{\cal{E}} , \quad [N_i, {\cal{E}}] =  i c {\cal{P}}_i
. \label{pary3}
\end{eqnarray}
-- commuting four momenta $[P_\mu, P_\nu]=0$\\The momentum
variables fulfill standard mass shell condition, described by mass
 Casimir
\begin{equation}
\label{pary4}
{\cal{E}}^2 - c^2 {\cal{P}}^2 \ = \ \mu^2 c^4 .
\end{equation}
Mass shall (3) is covariant under the
 relativistic boost transformations
\begin{equation}
\label{pary5} {\cal{E}}(\alpha) = {\cal{E}}\cosh(\alpha)
-c(\vec{n}\cdot\vec{\cal{P}})\sinh(\alpha)
.\end{equation}\begin{equation} \vec{\cal{P}}(\alpha) \ = \
\vec{\cal{P}} + \left((\cosh\alpha-1)\vec{n}\cdot\vec{\cal{P}} -
\frac{\cal{E}}{c}\sinh\alpha\right)\vec{n} .\end{equation} with
the rapidity--velocity relation $\vec{v} = c \vec{n} \tanh
\alpha\, ,\vec{\alpha}=\alpha\vec{n}$.\\ We shall consider invertible
nonlinear transformations of the momentum space in the form
\begin{equation}\label{pary6} \vec{\cal{P}}  = \vec{\cal{P}}(E,\vec{P}) 
=
 \vec{P} \ g \left(\frac{E}{\kappa c^2} ,
\frac{\vec{P}^{\,2}}{\kappa^2 c^2}\right)  , \ {\cal{E}} \ = {\cal
E}(E,\vec{P})=  \, \kappa c^2 f\left(\frac{E}{\kappa c^2} ,
\frac{\vec{P}^{\,2}}{\kappa^2 c^2}\right).
\end{equation}
with
  the dependence on
 dimensionfull mass-like parameter $\kappa$
satisfying the conditions
\begin{equation}
\lim_{\kappa\to\infty} g =  1 ,\qquad\qquad \lim_{\kappa\to\infty}
(\kappa c^2 f)  =  E .
\end{equation}
This form of transformations imply only changes in covariance
relations (\ref{pary3}) i.e. $[N , P]$ depend on functions $f$,
$g$ but the classical Lorentz algebra (\ref{pary2}) is not
changed.\\In transformed variables $E, \vec{P}$ the dispersion
relation (3) is given by
\begin{equation}
\kappa^2 c^4 f^2\left(\frac{E}{\kappa c^2} ,
\frac{\vec{P}^{\,2}}{\kappa^2 c^2}\right) - c^2 \vec{P}^{\,2}  g^2
\left(\frac{E}{\kappa c^2} , \frac{\vec{P}^{\,2}}{\kappa^2
c^2}\right) \ = \ inv. \ = \ \mu^2 c^4 .
\end{equation}
and it is invariant under nonlinear transformations of
boosts.
\\
For the special choices of functions $f$ and $g$ we obtain three
types of DSR and fourth class of SMSR theories, mentioned in the
Introduction.

\section{\bf 3. Doubly special relativity theories}

\subsection{DSR1 as nonlinear realization of Poincar\'{e}
algebra - an example}

We define nonlinear transformations of the momentum subalgebra 
as
follows
\begin{equation}
\vec{\cal{P}}  =  \vec{P} e^\frac{E}{\kappa c^2}\quad,\quad
{\cal{E}} =  \kappa c^2 \left(\sinh\frac{E}{\kappa c^2} +
\frac{\vec{P}^2}{2 \kappa^2 c^2}e^\frac{E}{\kappa c^2}\right) .
\end{equation}
In this new basis $(E, \vec{P})$ the covariance relations (\ref{pary3})
takes the form \begin{eqnarray} [M_i, P_j]& =  & i\epsilon_{ijk} P_k
,\qquad \qquad [M_i,E] =0 ,\cr [N_i,P_j]& =  & i \kappa c
\delta_{ij}\left[\sinh\left(\frac {E}{\kappa c^2}\right)
e^{-\frac{P_0}{\kappa c}}  + \frac1{2\kappa^2 c^2} (\vec P)^2\right] -
\frac {i}{\kappa c} P_i P_j , \cr [N_i, E ] & = &  i c P_i .
\end{eqnarray} DSR1 energy-momentum
 dispersion relation ($\kappa$-deformed mass Casimir) is given
by
\begin{equation}
C_2 \ = \ \left(2\kappa\sinh \frac{E}{2\kappa c^2} \right)^2 -
{1\over c^2}\vec P^2 e^{\frac{E}{\kappa c^2}}  = M^2 .
\end{equation}
Using the formulae (4-5) we get nonlinearly modified  boosts
transformations\footnote{The transformation (13) has been firstly
described for special choice of boost parameter $\vec{\alpha} =
(0,0,\alpha)$ in [6]; the general formula (13) was obtained
firstly in [2] and further discussed in [7]. }
\begin{equation}
E(\alpha)  =   E + \kappa c^2\ln W(\alpha, \vec{n}\vec{P}, E) .
\end{equation}
\begin{equation}
\vec{P}(\alpha) =  W^{-1}(...)\left[\vec{P} +
\left((\vec{n}\vec{P}) (\cosh\alpha - 1) - \kappa c B(m,E)
\sinh\alpha\right) \vec{n}\right] .\end{equation} where
\begin{eqnarray}
W(\alpha, \vec{n}\vec{P}, E) &= 1-\left(\frac{1}{\kappa c}
(\vec{n}\cdot\vec{P})\sinh\alpha + B(m,E) (1-\cosh\alpha)\right)
,\cr B(m,E) & =  1 - \cosh\left(\frac{m}{\kappa}\right)
e^{-\frac{E}{\kappa c^2}}
 = \frac{1}{2}\left(1 -
e^{-\frac{2E}{\kappa c^2}} + \frac{\vec{P}^{2}}{\kappa^2
c^2}\right) .
\end{eqnarray}

\subsection{DSR2 as nonlinear realization of Poincar\'{e}
algebra - an example}

We assume that the classical Lorentz algebra is given by the
formulae (\ref{pary2}). We define nonlinear transformations of the
momentum subalgebra as follows \begin{equation} \vec{\cal{P}}  = 
\vec{P}\left(1 - \frac{E}{\kappa c^2}\right)^{-1} ,\quad {\cal{E}} = 
E\left(1 - \frac{E}{\kappa c^2}\right)^{-1} . \end{equation} In this new
basis $(E, \vec{P})$ the covariance relations (\ref{pary3}) take the 
form
\begin{eqnarray} [M_i, P_j]& =  i\epsilon_{ijk} P_k , \qquad & 
[N_i,P_j] =
\frac{i}{c} \left(\delta_{ij} E - \frac{P_i P_j}{\kappa}\right) ,\cr
[M_i,E]& = 0 ,\quad & [N_i, E ] =  2i c \left(1 - \frac{E}{\kappa
c^2}\right) P_i . \end{eqnarray} DSR2 energy-momentum  
dispersion relation
is given by \begin{equation} C_2 = \frac{E^2 - c^2 \vec{P}^{\ 
2}}{\left(1
- \frac{E}{\kappa c^2}\right)^2} \ = \ M^2 c^4 . \end{equation} Using 
the
formulae (5-6) we get nonlinearly modified  boost transformations
\begin{equation} E(\alpha) = \left(E\cosh\alpha - c
(\vec{n}\vec{P})\sinh\alpha\right) {\cal{W}}^{-1}(\alpha, \vec{n}\vec{P},
E ) . \end{equation} \begin{equation} \vec{P}(\alpha)=  \left(\vec{P} +
\vec{n}\left((\cosh\alpha-1)\vec{n}\cdot\vec{P} -
\frac{E}{c}\sinh\alpha\right)\right) {\cal{W}}^{-1}(\alpha,
\vec{n}\vec{P}, E) .\end{equation} where \begin{equation}
{\cal{W}}(\alpha, \vec{n}\vec{P}, E)  =  1 + \frac{E}{\kappa c^2}
(\cosh\alpha - 1) - \frac{(\vec{n}\vec{P})}{\kappa c} \sinh\alpha
.\end{equation}

\subsection{DSR3 as nonlinear realization of Poincar\'{e}
algebra - an example}

We assume that the classical Lorentz algebra is given by the
formulae (\ref{pary2}). We define nonlinear transformations of the
 fourmomentum subalgebra as follows
\begin{equation}
\vec{\cal{P}}  =  \vec{P} ,\quad {\cal{E}} =  E \left(1 +
\frac{\vec{P}^2}{\kappa^2 c^2}\right)^{1/2} .
\end{equation}
In this new basis $(E, \vec{P})$ the covariance relations
(\ref{pary3}) takes the form
\begin{eqnarray}
[M_i, P_j]& =  i\epsilon_{ijk} P_k , \quad & [N_i,P_j] = {i\over
c} \delta_{ij} E \left(1 + \frac{\vec{P}^2}{\kappa^2
c^2}\right)^{1/2} ,\cr [M_i,E]& = [P_\mu, P_\nu] = 0 ,\quad  &
[N_i, E] = i c P_i \left(1 + \frac{\vec{P}^2}{\kappa^2
c^2}\right)^{-1/2}\left(1 - \frac{E^2}{\kappa^2 c^4}\right) .
\end{eqnarray}
The energy-momentum dispersion relation is given by
\begin{equation}
C_2 \ = \ E^2 \left(1 + \frac{\vec{P}^2}{\kappa^2 c^2}\right) -
c^2 \vec{P}^2 = M^2 c^4 .
\end{equation}
Then using the formulae (4-5) we get the  nonlinear boost
transformations
\begin{eqnarray}
E(\alpha) = \left(\frac{1+\vec{P}^2(\alpha)}{1 +
\vec{P}^2}\right)^{-{1\over 2}}\left[E\cosh \alpha -c
(\vec{n}\vec{P})\left(1 + \frac{\vec{P}^2}{\kappa^2
c^2}\right)^{-1/2} \sinh\alpha\right].\end{eqnarray}
\begin{equation}
\vec{P}(\alpha) \ = \ \vec{P} +
\left((\cosh\alpha-1)\vec{n}\cdot\vec{P} - \frac{E}{c}\left(1 +
\frac{\vec{P}^2}{\kappa^2 c^2}\right)^{1/2}
\sinh\alpha\right)\vec{n} .
\end{equation}

\section{SMSR as nonlinear realization of Poincar\'{e}
algebra - an example}

We assume that the classical Lorentz algebra is given by the
formulae (\ref{pary2}). We define nonlinear transformations of the
momentum subalgebra as follows
\begin{equation}
\vec{\cal{P}}  =  \vec{P} ,\quad {\cal{E}} =  2\kappa c^2
\sinh\left(\frac{E}{2\kappa c^2}\right) .
\end{equation}
In this new basis $(E, \vec{P})$ the covariance relations (\ref{pary3})
take the form \begin{eqnarray} [M_i, P_j]& =  i\epsilon_{ijk} P_k , 
\quad
&[N_i, E ]  =  i c P_i \cosh^{-1}\left(\frac{E}{2\kappa c^2}\right) ,\cr
[M_i,E]& = [P_\mu, P_\nu] = 0  , \quad &[N_i,P_j]  = 2i \kappa c
\delta_{ij} \sinh\left(\frac{E}{2 \kappa c^2}\right) . \end{eqnarray}
 dispersion relation is given by
\begin{equation}
C_2 \ = \ \left(2\kappa c^2\sinh \frac{E}{2\kappa c^2} \right)^2 -
c^2\vec P^2 = M^2 c^4 .
\end{equation}
Using the formulae (4-5) we get nonlinear  boost transformations
\begin{eqnarray}
E(\alpha) = 2\kappa c^2\mbox{arcsinh}
\left[\sinh\left(\frac{E}{2\kappa c^2}\right)\cosh \alpha -
\frac{(\vec{n}\vec{P})}{2\kappa c} \sinh \alpha\right].
\end{eqnarray}
\begin{equation}
\vec{P}(\alpha)  =  \vec{P} +
\left((\cosh\alpha-1)\vec{n}\cdot\vec{P} - 2\kappa c \sinh
\left(\frac{E}{2\kappa c^2}\right) \sinh\alpha\right)\vec{n} .
\end{equation}
One can see easily  that the values of $E(\alpha)$ and $P(\alpha)$
are not bounded.

\section{Final Remarks}

In this talk we mainly classified   different  nonlinear
bases for classical Poincar\'{e} algebra. If we observe that the
classical fourmomenta generators are endoved with primitive
coproducts $({\cal P}_\mu = ({\cal E}/c, {\cal P}_{i}))$
\begin{equation}\label{pary31}
  \Delta^{(0)}  {\cal P}_{\mu} = {\cal P}_\mu \otimes 1 + 1 \otimes
  {\cal P}_\mu\, .
\end{equation}
then by considering the formulas inverse to (\ref{pary6})
\begin{equation}\label{pary32}
E = \kappa \, c^2 F\left(\frac{\cal E}{\kappa c^2},
\frac{\vec{\cal P}^2}{\kappa^2
  c^2}\right)\, ,
  \qquad
  \vec{P} = \vec{\cal P} \, G \left(
  \frac{\cal E}{\kappa c^2}, \frac{\vec{\cal P}^2}{\kappa^2
  c^2}\right)\, .
\end{equation}
one obtains the symmetric nonlinear coproducts for the nonlinear
momentum $\vec{P}$ and nonlinear energy $E$.
 Such coproducts were considered in [2] as  describing the 
deformed
  addition law for fourmomenta
 and
further used in [8]  in the notation without the notion of coproduct.
 We get  the coproduct formulae:
\begin{equation}\label{pary33}
  \Delta { E} = \kappa \,  c^2 F \left(
  \frac{\Delta{\cal E}}{\kappa c^2},
  \frac{(\Delta \vec{\cal P})^2}{\kappa^2  c^2}\right)\, ,\quad
  \Delta \vec{P} = \Delta (\vec{\cal P}) G
  \left(
  \frac{\Delta {\cal E}}{\kappa c^2},
  \frac{(\Delta \vec{\cal P})^2}{\kappa^2 c^2},
  \right) .
\end{equation}
where
\begin{eqnarray}\label{pary34}
\Delta {\cal E}   = {\cal E} (E , \vec{P}) \otimes 1 + 1 \otimes
{\cal E} (E, \vec{P}) ,\cr \Delta \vec{\cal P} = \vec{\cal P} (E,
\vec{P})\otimes 1 + 1 \otimes \vec{\cal P}(E,\vec{P}) .
\end{eqnarray}
The quantum nonsymmetric
  coproduct is obtained if we modify primitive coproduct
(\ref{pary31}) by introducing Drinfeld twist $T$ [1]
\begin{equation}\label{pary35}
\Delta {\cal P}_{\mu} = T^{-1} \circ \Delta^{(0)} ({\cal P}_\mu)
\circ T = \Delta^{(0)} ({\cal P}_\mu) + [ r, {\cal P}_\mu] +
\ldots .
\end{equation}
where $r$ is the classical Poincar\'{e} $r$-matrix [9]. In such a
way we obtain the quantum Poincar\'{e} algebras with nonlinear
fourmomentum basis  given by (32) and nonprimitive coproduct (33)
 with inserted formulae (35).

 \end{document}